# Hilbert curves in 2 dimensions generated by L-systems

Arie Bos [1]

**Abstract:**

*Generating Hilbert curves in $\mathbb{Z}^2$ using L-systems appears to be efficient and easy.*

## 1. Introduction

***N.B. 1.*** Apart from the organization of this paper in sections, most of the text consists of statements which are divided in (numbered) Definitions, Examples, Theorems and Notes, the last ones also numbered and indicated by N.B. This makes cross references within the text much easier.

***N.B. 2.*** In [1] Hilbert presented the later so called Hilbert curves:

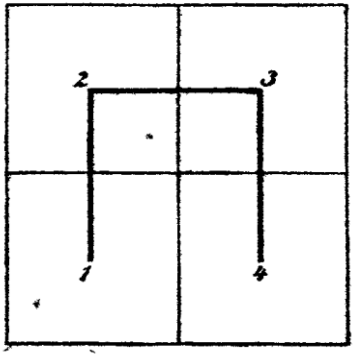 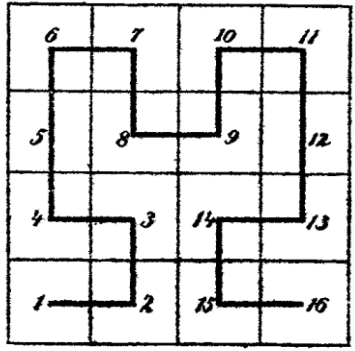 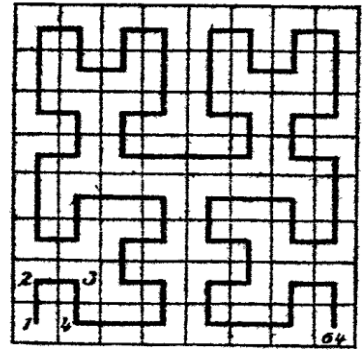

Fig. 1.  Fig. 2.  Fig. 3.

In this paper a method is presented to generate this curve easily and fast by L-systems.

***N.B. 3.*** To represent these curves in an easy way, we adopt the following terminology: a horizontal line of one unit to the right is indicated by a 1, one unit horizontal to the left by -1 and the same for 2 in the vertical direction: 2 represents one unit up and -2 one unit down. So following the order given by Hilbert in his original paper, $2, 1, -2$ and

$1, 2, -1, 2, 2, 1, -2, 1, 2, 1, -2, -2, -1, -2, 1$ are the representations of the first two curves.

***N.B. 4.*** By rotating the odd Hilbert curves over 90° we can arrange the representations in such a way that each curve is the beginning of a next one. The first and third then look like

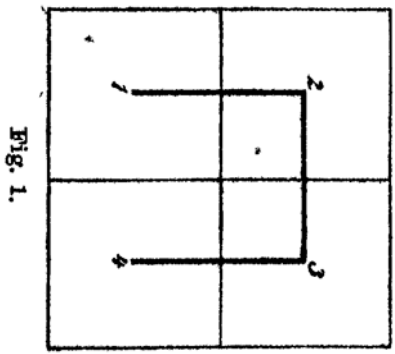 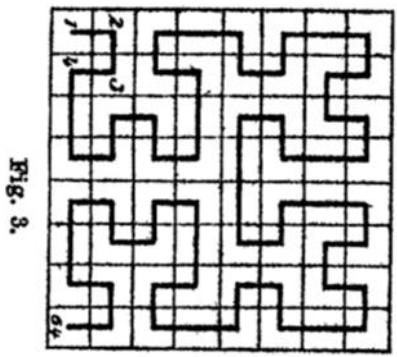

---

[1] arie.bos@planet.nl

*N.B. 5.* First I will introduce L-systems which is shorthand for Lindenmayer systems. After that an L-system generating the Hilbert curve(s) will be represented.

## 2. L-system

**Definition 1.** An L-system [2] consists of V, an *alphabet* (of symbols),
$V^* = V \cup (V \times V) \cup (V \times V \times V) \cup \cdots = \bigcup_{k \geq 1} V^k$ the set of all *strings* (words) of V (with * being the Kleene star) and *production rules*, a mapping $P: V \to V^*$.

*N.B. 6.* P can be extended to $P: V^* \to V^*$, where the same letter $P$ is used for the extended mapping, by $P(x_1, x_2, x_3, \cdots) = P(x_1), P(x_2), P(x_3), \cdots$ where $x_i \in V$ $(i \geq 1)$ and the strings are concatenated. (Since a concatenation of strings is again a string, the mapping $P$ can be applied 0 or more times, denoted by $P^n$ for some $n \geq 0$.)

*N.B. 7.* $\{x \in V \mid P(x) = x\}$ are *constants* and $\{x \in V \mid P(x) \neq x\}$ are *variables*. For a special $s \in V^*$, mostly $s \in V$, called the *start*, *generation n* is $P^n(s)$; the set of all generations is $\{P^n(s) \mid n \geq 0\}$

*N.B. 8.* L-systems are perfectly equipped for *repeated squaring*, see the next section.

**Example 1.** The 2-dimensional Hilbert curves are generated by the L-system with $V = \{0, 1, -1, 2, -2, 3, -3, 4, -4\}$, 0 being the start element and the production rules being

$$P = \begin{cases} 0 \to 0, 3, 4, -1 \\ 1 \to 2, 3, 4, -1 \\ 2 \to 1, 3, 4, -1 \\ 3 \to 4, 2, 1, -4 \\ 4 \to 3, 2, 1, -4 \end{cases} \text{ and } P(-x) = -P(x)$$

*N.B. 9.* In transforming a string like $0, 3, 4, -1, 4, 2, 1, -4, 3, 2, 1, -4, -2, -3, -4, 1$ to a graph the following prescription is adopted: 0 makes the graph start in the origin, 3/-3 and 4/-4 are replaced by 1/-1 and 2/-2 respectively and a 1 means a horizontal line of one unit to the right, -1 one unit horizontal to the left and the same for 2 in the vertical direction: 2 equals one unit up and -2 equals one unit down. So this string gives the second (rotated) Hilbert curve as one can easily check. So the string which represents the 2$^{nd}$ Hilbert curve is $1, 2, -1, 2, 2, 1, -2, 1, 2, 1, -2, -2, -1, -2, 1$. The ultimate Hilbert string is not (yet) in the OEIS [3].

*N.B. 10.* An alternative L-system for generating the Hilbert curve is

$$V = \{0, 1, -1, 2, -2, 3, -3, 4, -4\} \text{ and } P = \begin{cases} 1 \to 1 \\ 2 \to 2 \\ 3 \to 4, 2, 3, 1, 3, -2, -4 \\ 4 \to 3, 1, 4, 2, 4, -1, -3 \end{cases} \text{ and } P(-x) = -P(x), \text{ 3 being the}$$

start element. In translating a string from the L-system to a curve, ignore the 3/-3 and 4/-4 and interpret 1/-1 and 2/-2 as usual. Notice that here the L-system describes how we explode a point in the curve to a piece of the curve on a next level.

*N.B. 11.* Other L-systems which give the same result are possible, then often a different notation is used, not seldom to be applied by turtle graphics [4].

## 3. The Hilbert curves

***N.B. 12.*** Let us calculate $P(0)$, $P^2(0)$ and $P^3(0)$ from Example 1. We get,

$P(0) = 0, 3, 4, -1 = 1, 2, -1$

$P^2(0) = 0, 3, 4, -1, 4, 2, 1, -4, 3, 2, 1, -4, -2, -3, -4, 1 = 1, 2, -1, 2, 2, 1, -2, 1, 2, 1, -2, -2, -1, -2, 1$ and,

with 3 and 4 replaced :

$P^3(0) = 1, 2, -1, 2, 2, 1, -2, 1, 2, 1, -2, -2, -1, -2, 1, 1, 2, 1, -2, 1, 1, 2, -1, 2, 1, 2, -1, -1, -2, -1, 2, 2, 2, 1, -2,$
$\qquad 1, 1, 2, -1, 2, 1, 2, -1, -1, -2, -1, 2, -1, -1, -2, 1, -2, -2, -1, 2, -1, -2, -1, 2, 2, 1, 2, -1$

which gives rise to the following graphs.[2] As we can see each graph is the starting figure for the next.

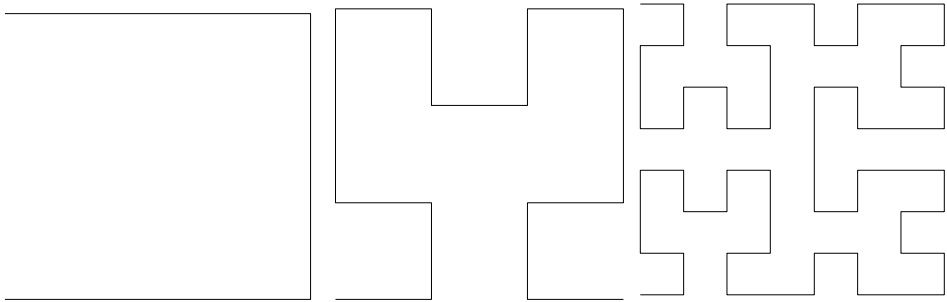

***N.B. 13.*** By induction it is obvious that the $m^{th}$ Hilbert curve has $2^{2m} - 1$ sides and $2^{2m}$ nodes.

***N.B. 14.*** By multiplying the $m^{th}$ curve by $\sqrt{2^{-m}}$ we get a sequence of curves with a space filling limit.

***N.B. 15.*** From N.B. 10 we get $P^n(3) = (P^n(4), 2, P^n(3), 1, P^n(3), -2, -P^n(4))$ which is just an analogue of the construction in [5].

## 4. Repeated squaring

***N.B. 16.*** If we calculate $P^k$ we mostly do it iteratively by $P(P^{k-1})$, which is linear since it requires applying $P$ $k$ times. But with a well-known technique this can be done much more efficiently. I'll give an example to illustrate it.

**Example 2.** Let us calculate $P^{100}$. Notice that $100 = (1\ 1\ 0\ 0\ 1\ 0\ 0)_2$, so $100 = 2^6 + 2^5 + 2^2$ which leaves us with calculating $P^m, m = 2^6$ by subsequent squaring

$P \longrightarrow P^2 \longrightarrow P^4 \longrightarrow P^8 \longrightarrow P^{16} \longrightarrow P^{32} \longrightarrow P^{64}$. Notice that we have $P^{32}$ and $P^4$ as well. By multiplying these 3 we get $P^4 \times P^{32} \times P^{64} = P^{100}$ and the whole process took $6 + 3 = 9$ multiplications, compared to the 99 which a linear approach would have taken.

## 5. Conclusions

***N.B. 17.*** Representing a Hilbert curve by a sequence of numbers from $1, -1, 2, -2$ and using an L-system to generate these sequence(s), we have a very concise and fast method of constructing the Hilbert curve(s).

***N.B. 18.*** In a next paper we will set up a more general theory to generate all kinds of fractal curves in a fast and compact way with the aid of L-systems.

---

[2] The calculations and the last graphs were made with the array programming language J [6]